\documentclass[aps,twocolumn,a4paper,pra,showpacs]{revtex4}%
\usepackage{psfig}
\usepackage{amssymb}
\usepackage{amsmath}
\usepackage{amsfonts}
\usepackage{graphicx}%
\begin{document}
\title{Casimir effect with rough metallic mirrors}
\author{Paulo A. Maia Neto}
\affiliation{Instituto de F\'{\i}sica, UFRJ, Caixa Postal 68528, 21945-970 Rio de Janeiro RJ, Brazil}
\author{Astrid Lambrecht}
\author{Serge Reynaud}
\affiliation{Laboratoire Kastler Brossel,
CNRS, ENS, UPMC case 74, Campus Jussieu, 75252 Paris, France}

\begin{abstract}
We calculate the second order roughness correction to the Casimir energy for two parallel metallic mirrors. 
Our results may also be applied to the plane-sphere geometry used in most experiments. 
The metallic mirrors are described by the plasma model,
with arbitrary values for the plasma wavelength, the mirror separation and the roughness
correlation length, with the roughness amplitude remaining the smallest length scale for perturbation theory to hold.
 From the analysis of the intracavity field fluctuations, 
 we obtain the Casimir energy correction in terms
of generalized reflection operators, which account for diffraction and polarization coupling
in the scattering by the rough surfaces. We present simple analytical expressions for several limiting cases,
 as well as numerical results that allow for a reliable
calculation of the roughness correction in real experiments. 
The correction is larger than the result of the 
Proximity Force Approximation, which
is obtained from our theory as a limiting case (very smooth surfaces). 
\end{abstract} 


\maketitle

\section{Introduction}

The Casimir force of attraction between metallic mirrors~\cite{Casimir}
has been measured with high experimental precision over the last 
few years~\cite{experimental}. These new 
experiments allow for an accurate theory/experiment comparison 
\cite{theory_exp}, opening the way for the search for new weak forces 
with submillimetric ranges \cite{search_nwf}.
On the theoretical front, accurate results based on realistic models 
are sorely needed in order to match the desired levels of accuracy. 
Three important effects provide the main corrections to the ideal 
configuration considered by Casimir: non zero temperature~\cite{Genet00}, 
finite conductivity~\cite{Heinrichs,Lambrecht2000} and
roughness of the mirrors \cite{Bree}-\cite{EPL}. Temperature corrections are important 
 when the distance $L$ between the 
mirrors is  above $1 \mu{\rm m},$ whereas finite conductivity and roughness
provide the major corrections for the short distances (of the order of a few hundred
nanometers) probed by most experiments. 

In principle, these effects must be taken into account simultaneously. The overall 
correction is not in general the product of the separate corrections calculated 
independently. In particular, the correlation between finite conductivity and roughness effects 
is essential, because they
both intervene at the same range of $L.$ Therefore, a reliable 
theory for short distances must analyze the roughness 
effect in the context of a finite-conductivity model for the material medium. To this aim, 
we describe the optical properties of the metallic mirrors by the plasma model. 

When the surface profiles are nearly smooth over distances of the order of $L,$
the roughness correction may be calculated from the 
Proximity Force Approximation (PFA)~\cite{PFA}. 
In this approximation, the 
Casimir energy is computed from the 
 formula for parallel planes by averaging the 
`local' distance over the surface~\cite{EPL}. 
In order to derive more general results, we 
develop a perturbative
theory for the Casimir energy with rough plane
mirrors, allowing for the computation of the energy correction when the 
surface profile varies on arbitrarily short length scales, provided that
they are larger than the  roughness amplitude (otherwise the perturbative 
approximation would not apply).  
Our approach is  applicable to 
most  Casimir force measurements between metallic mirrors. 

We follow the approach of Ref.~\cite{lossy_cavities}, and 
consider the two mirrors as a plane Fabry-Perot cavity, which is treated as composed optical 
network in order to calculate the  intracavity
field fluctuations.   
We then derive a formal result for
the Casimir energy up to
second order in the amplitude of roughness,
in terms of generalized reflection coefficients describing the 
scattering by rough surfaces, taking into account the coupling between 
Transverse Electric (TE) and Transverse Magnetic (TM) polarizations. 
Numerical results derived from the present calculation were presented in a letter~\cite{letter2},
together with  some analytical limiting cases. 
In this paper we present the complete derivation and the explicit formulas 
used in  \cite{letter2}.

This paper is organized in the following way. In Sec.~II, we 
present some basic definitions and assumptions, and discuss
the validity of the PFA in two different contexts. 
In Sec.~III, we
derive the formal, general result 
for the second order roughness energy correction,
 which is then applied to the specific plasma-model calculation presented
in Sec.~IV. Several limiting cases are discussed in the 
following sections: the short roughness wavelength regime (Sec.~V), 
the perfectly-reflecting limit (Sec.~VI) and the plasmon limit 
(Sec.~VII). In Sec.~VIII, we discuss the example 
of a Gaussian roughness spectrum 
and present some concluding remarks.
Three appendices present additional details of the derivations.

\section{General considerations and assumptions}

Our Fabry-Perot cavity of length $L$ is composed of two parallel mirrors with rough surfaces, 
as shown in Fig.~1. We analyze the cavity as a composed optical network, and
calculate the  fluctuations of the intracavity fields propagating along the positive and negative 
$z$-axis,
$\stackrel{\rightarrow}{\rm E}_{\rm C}$ and 
$\stackrel{\leftarrow}{\rm E}_{\rm C}$, 
in terms of the fluctuations of the incoming free-space fields ${\rm E}_{\rm L}^{\rm in}$ and 
${\rm E}_{\rm R}^{\rm in}$   (also shown in Fig.~1 are the outgoing fields   ${\rm E}_{\rm L}^{\rm out}$ and 
${\rm E}_{\rm R}^{\rm out}$). In appendix A, we show that 
the Casimir energy turns out to depend only 
on the coefficients describing 
the reflection of the intracavity fields
by the internal sides of mirrors M1 and M2. 
The  functions 
$h_1\left(\mathbf{r}\right)$
and $h_2\left(\mathbf{r}\right)$
define their surface profiles  with respect to reference planes at $z=0$ (see Fig. 2) and 
$z=L,$ respectively.
$\mathbf{r}$ collects the two
transverse coordinates $\left( x,y\right)$ orthogonal to the cavity  extension. 
By construction, both $h_1$ and $h_2$ have zero
spatial averages: $\langle h_j \rangle =0,j=1,2,$ and are
counted as positive when they correspond to local length decreases below the mean value $L$. 

\begin{figure}[ptb]
\centerline{\psfig{figure=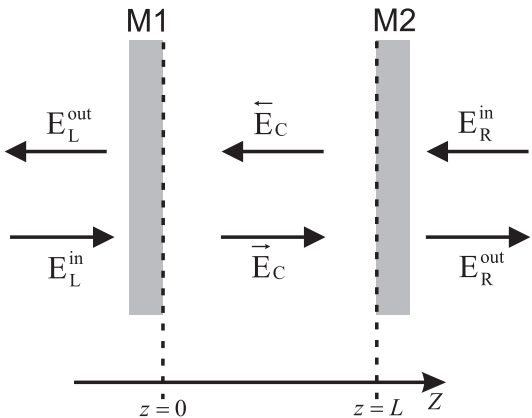,width=7cm}}\caption{Fabry-Perot cavity of length $L$.}
\end{figure}

We assume that the two surfaces  are statistically independent, so that the cross correlation function vanishes:
\begin{equation}
\langle h_1({\bf r}) h_2({\bf r'}) \rangle  = 0.
\end{equation}
Translational symmetry on the $xy$ plane
implies that the self 
correlation functions satisfy
\[
\langle h_j({\bf r}) h_j({\bf r'}) \rangle  = \langle h_j({\bf r-r'}) h_j({\bf 0}) \rangle, \;\;\; j=1,2.
\]
Then in the Fourier domain we have (${\bf k}$ is a two-dimensional vector)
\begin{equation}\label{translational1}
\langle H_j({\bf k})H_j({\bf k}') \rangle = (2\pi)^2 \, \delta^{(2)}({\bf k}+{\bf k}')\, \sigma_{jj}({\bf k}),
\end{equation}
where $H_j({\bf k})$ is the Fourier transformation of $ h_j({\bf r})$  and the roughness spectrum $\sigma_{jj}({\bf k})$ is the Fourier transform of the self
correlation function: 
\[
\sigma_{jj}({\bf k})=\int d^2{\bf r}\,{\rm e}^{-i{\bf k}\cdot{\bf r}}
\langle h_j({\bf r}) h_j({\bf 0}) \rangle.
\]

We assume that the area $A$ of the mirrors contains many 
correlation areas: $A \gg \ell_C^2$, where $\ell_C$ is the correlation length characteristic of
the self correlation function. In this case, a single mirror already contains many
independent realizations of surface profiles, and hence spatial and ensemble averages are 
equivalent.

\begin{figure}[ptb]
\centerline{\psfig{figure=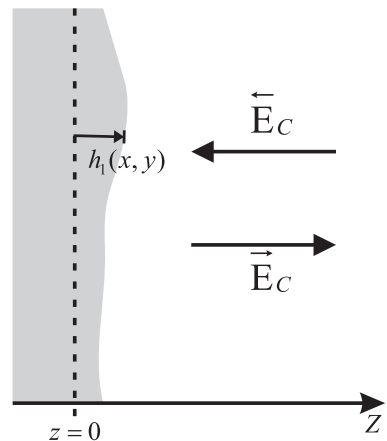,width=5cm}}\caption{Magnified detail of the internal surface of  mirror M1.}
\end{figure}

We also assume that the deformation amplitudes
are very small, in the scale of the mean cavity length, $|h_j|\ll L,$
as well as in the scale of the correlation length,  $|h_j|\ll \ell_C,$
so that the surface profile gradients satisfy $|{\bf \nabla} h_j|\ll 1.$
This allows us to treat the surface deformations as  small perturbations of the ideal
plane geometry. The Casimir energy is then calculated up to second order of the 
deformation amplitudes. 
More precisely, the energy correction is obtained in terms of 
\begin{equation}
\langle H_j({\bf k})H_j(-{\bf k}) \rangle = 
 A\sigma_{jj}({\bf k}), \label{area2}
\end{equation}
where we have used Eq.~(\ref{translational1}).

Most  experiments are performed with a plane-sphere (PS) setup, instead of 
the plane-plane (PP) cavity used as the benchmark for our perturbative calculation. 
However,  our results may also be applied to those experiments, 
provided that we use the PFA to connect the two different geometries.
 In this case, the force $F_{\rm PS}$ between a sphere of radius $R$ and 
a plane at a distance of closest approach $L$ 
is given in terms of the energy $E_{\rm PP}$ for the plane-plane cavity as follows: 
\begin{equation}
F_{\mathrm{PS}}\left(  L\right)  =2\pi R\frac{E_{\mathrm{PP}}\left(  L\right)
}{A} \label{FPSsmooth}
\end{equation}
The relative roughness correction of the force in the plane-sphere geometry may then be obtained from 
the relative energy correction calculated in this paper:
\begin{equation}\label{PFAsphere}
\Delta=\frac{\delta F_{\rm PS}}{F_{\rm PS}}= \frac{\delta E_{\rm PP}}{E_{\rm PP}}.
\end{equation}
We emphasize that the PFA amounts to the addition of contributions
corresponding to different local inter-plate distances, assuming these contributions to be
independent. But the Casimir energy is not additive, so that the PFA cannot be
exact, although it is often improperly called a theorem. 

Note that the conditions required for applying the PFA 
to connect the plane-sphere and the plane-plane geometries 
are quite different from those necessary for using the PFA in the computation of 
the roughness correction itself. In the first 
case, it is necessary that  the radius $R$  is large enough, so that the 
separation $L$ satisfies $L\ll R$ 
~\cite{PFAold}. 
Moreover, to avoid any interplay between 
curvature and roughness effects, one requires  the  correlation length $\ell_C$
 to be small enough, so that many correlation 
areas are contained in a given nearly-plane local section of the spherical surface: $\ell_C^2\ll RL.$
In contrast, applying the PFA to roughness requires the surfaces to be nearly plane in the 
scale of the separation: $\ell_C \gg L.$ 
Then, the following second-order roughness correction to the Casimir energy is obtained~\cite{EPL}: 
\begin{equation}
\delta {  E}_{\rm PP} \approx \frac{E_{\rm PP}''(L)}{2}\langle h_1^2 + h_2^2 \rangle,\;\;\;\;\;\;\;\;\;\;{\rm (PFA)}\label{PFA1}
\end{equation}
Hence, the PFA result for the roughness correction depends
on the second order derivative  of the energy and on the variances of the length deformations $h_1$ and $h_2$.
This expression is equivalent to the procedure used for analyzing the effect
of roughness in recent experiments \cite{Harris00,Klimchitskaya99}.
In the following sections, we will assume the PFA 
to provide a valid description of curvature, but not of roughness.

\section{Reflection and loop functions for rough mirrors}

In this section we start to 
develop a description of  reflection by the internal sides of the cavity mirrors (see Fig.~1), 
leaving the more general theory which takes into account the coupling 
with the external fields to appendix A. 

We take the mixed Fourier representation for the intracavity fields:
\begin{equation}\label{incident}
\stackrel{\rightarrow}{\bf E}_{\rm C}({\bf k},z,\omega)= 
(\stackrel{\rightarrow}{\rm E}_{\rm C}^{\scriptscriptstyle\rm TE}({\bf k},\omega)
{\hat \epsilon}^{\scriptscriptstyle\rm TE}
+\stackrel{\rightarrow}{\rm E}_{\rm C}^{\scriptscriptstyle\rm TM}({\bf k},\omega){\hat \epsilon}^{\scriptscriptstyle\rm TM})\,e^{ik_z z},
\end{equation}
where $\omega$ is the frequency,  ${\bf k}$ is the two-dimensional wavevector associated to 
propagation parallel to the $xy$ plane, and
$k_z = {\rm sgn}(\omega)\sqrt{\omega^2/c^2-k^2}$, with ${\rm sgn}$ denoting the sign function. 
The complete wavevector is given by ${\bf K}={\bf k}+ k_z {\hat z}.$
 The  field $\stackrel{\leftarrow}{\bf E}_{\rm C}$
is written in a similar way, except for the 
replacement $k_z\rightarrow -k_z.$ 
The  TE and TM unitary vectors are defined  in the following way:
\begin{equation}
{\hat \epsilon}^{\scriptscriptstyle\rm TE}= {\hat z}\times{\hat k},
\end{equation}  
\begin{equation}
{\hat \epsilon}^{\scriptscriptstyle\rm TM}= {\hat \epsilon}^{\scriptscriptstyle\rm TE}\times{\hat K}.
\end{equation}  

It is useful to employ the Dirac notation,
with the ket $|\stackrel{\rightarrow}{\rm E}_{\rm C}(\omega)\rangle$
providing a compact notation for the field amplitudes:
\[
\stackrel{\rightarrow}{\rm E}_{\rm C}^{p}({\bf k},\omega) = 
\langle {\bf k},p|\stackrel{\rightarrow}{\rm E}_{\rm C}(\omega)\rangle,
\]
where $p={\rm TE},{\rm TM}$ denotes the polarization.

The  reflection by the mirror M1 at frequency $\omega$ is written as  
\begin{equation}\label{reflection-operator}
|\stackrel{\rightarrow}{\rm E}_{\rm C}(\omega)\rangle = {\cal R}_1(\omega)
|\stackrel{\leftarrow}{\rm E}_{\rm C}(\omega)\rangle.
\end{equation}
When the surface is rough, the  operator ${\cal R}_1(\omega)$ mixes up different
values of ${\bf k}$ and
polarizations. On the other hand, the field frequency is conserved, since the surface is at rest.
The explicit form of (\ref{reflection-operator}) is 
\begin{equation}\label{reflection-operator-explicit}
\stackrel{\rightarrow}{\rm E}_{\rm C}^{p}({\bf k},\omega)=
\int\frac{d^2{\bf k}'}{(2\pi)^2}\sum_{p'}\langle {\bf k},p|{\cal R}_1(\omega)|  {\bf k'},p'\rangle \, 
\stackrel{\leftarrow}{\rm E}_{\rm C}^{p'}({\bf k}',\omega).
\end{equation}
The reflection by  mirror M2 is defined in a similar way in terms of the operator 
${\cal R}_2(\omega).$ 

We expand the reflection operators ${\cal R}_j(\omega),$ $j=1,2,$ 
in powers of the deformation amplitudes $h_j:$ 
\begin{equation}
{\cal R}_j(\omega) =  {\cal R}_j^{(0)}(\omega) + \delta{\cal R}_j^{(1)}(\omega) + \delta{\cal R}_j^{(2)}(\omega).
\end{equation}
The zero-th order operators ${\cal R}_j^{(0)}$
correspond to ideally plane surfaces. They do not modify the polarization nor the 
momentum ${\bf k},$ and hence  
 are diagonal in the basis 
$\{ |{\bf k},p\rangle \}:$
\begin{equation}\label{reflection-plane}
\langle {\bf k},p|\, {\cal R}_j^{(0)}(\omega)\, |  {\bf k'},p'\rangle = 
(2\pi)^2\delta^{(2)}({\bf k}-{\bf k}')\,\delta_{pp'}\, r^p_{j}({\bf k},\omega),
\end{equation}
where $r^p_{j}({\bf k},\omega)$ are the specular reflection coefficients for a plane mirror.

For the ideal Fabry-Perot cavity, the Casimir effect may be entirely described 
by these reflection coefficients, which characterize the optical properties of the cavity as 
seen by the intracavity field~\cite{jaekel91}.
As shown in Appendix A,
a similar result holds for a cavity with rough mirrors, except that 
the specular reflection coefficients are replaced by the reflection operators defined above.
The Casimir force is calculated from the spectral density characterizing 
the vacuum field fluctuations. 
For the intracavity field, the free-space spectral density 
for polarization $p$ is multiplied by the generalized Airy function
$g_p({\bf k}, \omega),$ 
which quantifies the joint boundary effect of the two mirrors. 
 We then derive the Casimir force  after
including the contribution of evanescent waves:
\begin{equation}\label{Fcas}
F_{\rm PP} = A\sum_p
 \int\frac{d^2 k}{(2\pi)^2}
\int_0^{\infty}\frac{d \omega}{2\pi}
\hbar k_z
(1-g_p({\bf k}, \omega)).
\end{equation}
$F_{\rm PP}$ is defined as the $z$-component of the force on mirror M1; hence it is positive
in case of attraction. 

We compute $g_p({\bf k}, \omega)$ up to second order in $h_1$ and $h_2:$
\begin{equation}
\label{exp_g}
g_p({\bf k}, \omega) = g_p^{(0)}({\bf k}, \omega)+\delta g_p^{(1)}({\bf k}, \omega)+
\delta g_p^{(2)}({\bf k}, \omega).
\end{equation}  
$g_p^{(0)}({\bf k}, \omega)$ is the Airy function for the ideal plane cavity~\cite{lossy_cavities}: 
\[
 g_p^{(0)}({\bf k}, \omega)= 1 + f_p({\bf k},\omega)+f_p({\bf k},\omega)^*,
\]
where $f_p({\bf k},\omega)$ is the corresponding loop function.
It is given by the superposition of all propagation factors representing
a closed loop with $n$ round-trips inside the cavity:  
\begin{widetext}
\begin{equation}
f_p({\bf k},\omega)=\sum_{n=1}^{\infty}(r^p_1({\bf k},\omega)r^p_2({\bf k},\omega)e^{-2\kappa L})^n=
\frac{r^p_{1}({\bf k},\omega)r^p_{2}({\bf k},\omega)e^{-2\kappa L}}{1-r^p_{1}({\bf k},\omega)
r^p_{2}({\bf k},\omega)e^{-2\kappa L}},\label{fp}
\end{equation}
\end{widetext}
where 
$\kappa = - i \sqrt{\omega^2/c^2-k^2}.$  
When replacing 
$ g_p$ by
$ g_p^{(0)}$ in (\ref{Fcas}), we find the well-known result for the Casimir
 force in the ideal case~\cite{jaekel91}. 

The first-order Casimir force correction, coming from 
$\delta g_p^{(1)}({\bf k}, \omega)$ in (\ref{exp_g}), vanishes because it is proportional to
the averages $\langle h_1 \rangle$ and   $\langle h_2 \rangle.$ 
Thus, the roughness correction is of second order, and results from the contribution of 
$\delta g_p^{(2)}({\bf k}, \omega).$
These functions are written in terms of  `rough' loop functions
$\delta f^{\rm (2\,i)}_p({\bf k},\omega)$ and $\delta f^{\rm (2\,ii)}_p({\bf k},\omega),$
 gathering the second-order
contributions of the first ($\delta{\cal R}_j^{(1)}$) and second order 
($\delta{\cal R}_j^{(2)}$) reflection operators, respectively:    
\begin{equation}\label{soma}
\delta g_p^{(2)}({\bf k}, \omega) =  \delta f^{(2\,\rm i)}_p({\bf k},\omega)+ 
\delta f^{(2\,\rm ii)}_p({\bf k},\omega)+{\rm c.c.}.
\end{equation}

$\delta f^{\rm (2\,ii)}_p({\bf k},\omega)$ is the superposition of all closed loops involving a single 
second-order rough reflection at one of the mirrors:
\begin{widetext}
\begin{equation}\label{fii}
\delta f^{(2\,\rm ii)}_p({\bf k},\omega)=\frac{1}{A}\sum_{j=1}^2\langle{\bf k},p| {\cal D}(\omega)^{-1}
\,e^{-{\cal K}(\omega) L}\,
\mathcal{R}_{[j+1]}^{(0)}(\omega)\,
e^{-{\cal K}(\omega) L}\,
\delta\mathcal{R}_{j}^{(2)}(\omega)\,{\cal D}(\omega)^{-1}
| {\bf k}, p \rangle,
\end{equation}
\end{widetext}
with $[j+1]$ representing a sum modulo 2.  
Like ${\cal R}_j^{(0)}(\omega)$
in Eq.~(\ref{reflection-plane}),  the operators 
${\cal D}(\omega)$ and ${\cal K}(\omega)$ are diagonal operators,  with elements 
$1-r^p_{1}({\bf k},\omega)r^p_{2}({\bf k},\omega)e^{-2\kappa L}$
and $\kappa,$ respectively. 

To understand why  $\delta f^{(2\,\rm ii)}_p({\bf k},\omega)$  
is a generalization of the ideal loop function $f_p({\bf k},\omega),$
we should read the r.-h.-s. of Eq.~(\ref{fii}) 
from right to left. The second-order rough reflection at 
mirror $j$ is followed by a one-way propagation between the two mirrors
(operator $\exp(-{\cal K}(\omega) L)$), and  then by a specular reflection 
at mirror $[j+1].$ The loop is closed by a second one-way propagation back to mirror $j.$ 
This loop 
can be preceeded and/or followed by arbitrary numbers of 
round-trips with specular reflections, 
hence the entire expression is sandwiched between two operators 
${\cal D}(\omega)^{-1}.$ 

`Closing the loop' means 
to ensure that the initial and final states
are the same, which is represented by
the ket $| {\bf k}, p \rangle$ and the corresponding bra in Eq.~(\ref{fii}). 
Since all zero-th order processes 
conserve momentum and polarization, only second-order rough reflections 
that also conserve momentum and polarization are allowed,
so that only {\it diagonal} elements of $\delta\mathcal{R}_{j}^{(2)}$ 
are expected to contribute in Eq.~(\ref{fii}). Its 
explicit evaluation indeed yields~\cite{foot1}
\begin{equation}\label{fii_explicit}
\delta f^{\rm (2\,ii)}_p({\bf k},\omega)=\frac{1}{A}\sum_{j=1}^2\frac{ {r}^{p}_{[j+1]}({\bf k},\omega)
\langle {\bf k},p| \delta{\cal R}_j^{(2)}(\omega)| {\bf k}, p \rangle e^{-2\kappa L}}
{(1-r^p_{1}({\bf k},\omega)r^p_{2}({\bf k},\omega)e^{-2\kappa L})^2}.
\end{equation}
Those diagonal matrix elements are of the form
\begin{widetext}
\begin{equation}\label{R2}
\langle {\bf k},p| \delta{\cal R}_j^{(2)}(\omega)| {\bf k}, p \rangle = \int \frac{d^2k'}{(2\pi)^2} 
R_{j;p}^{(2)}({\bf k},{\bf k}';\omega)\,
|H_j({\bf k}-{\bf k}')|^2,
\end{equation}
\end{widetext}
where the  non-specular coefficients $ R_{j;p}^{(2)}({\bf k},{\bf k}';\omega)$
are independent of the profile functions $H_j({\bf k}).$ 

On the other hand,  nondiagonal matrix elements of $\delta{\cal R}_j^{(1)}$ contribute 
to the loop function $\delta f^{\rm (2\,i)}_p({\bf k},\omega),$
because the latter contains 
two first-order rough reflections instead of just one second-order reflection.
These elements are 
 of the form
\begin{equation}\label{R1}
\langle {\bf k}, p| \delta{\cal R}_j^{(1)}(\omega)| {\bf k'}, p' \rangle=
R_{j;pp'}^{(1)}({\bf k},{\bf k}';\omega)\,H_j({\bf k}-{\bf k}'),
\end{equation}
where again the coefficients $R_{j;pp'}^{(1)}({\bf k},{\bf k}';\omega)$ are independent 
of  $H_j({\bf k}).$ According to this expression, a given Fourier component $\Delta {\bf k}$ 
of the surface profile leads to a field momentum modification by $\Delta {\bf k}.$

Since the elements 
$\langle {\bf k}, p| \delta{\cal R}_j^{(1)}(\omega)| {\bf k'}, p' \rangle$ are proportional to 
$H_j({\bf k}-{\bf k}')$, 
the terms associated to first-order rough reflections at {\it different} mirrors are proportional either to
the product 
\[
H_1({\bf k}-{\bf k}')H_2({\bf k}'-{\bf k})
\]
or to its complex conjugate. 
As discussed in Sec.~II, we assume that their average values vanish because 
the two  surface profiles are statistically independent. 
Thus, we only
 keep  the terms associated to two first-order rough reflections at the same
mirror  when deriving $\delta f^{\rm (2\,i)}_p({\bf k},\omega)$ (omitting the dependence with $\omega$
in the r.-h.-s.):
\begin{widetext}
\begin{equation}\label{fi}
\delta f^{(2\,\rm i)}_p({\bf k},\omega)=\frac{1}{A}\sum_{j=1}^2\langle{\bf k},p| {\cal D}^{-1}
\,e^{-{\cal K} L}
\,\mathcal{R}_{[j+1]}^{(0)}
\,e^{-{\cal K} L}
\,\delta\mathcal{R}_{j}^{(1)}
\,{\cal D}^{-1}
\,e^{-{\cal K} L}
\,\mathcal{R}_{[j+1]}^{(0)}
\,e^{-{\cal K} L}
\,\delta\mathcal{R}_{j}^{(1)}\,{\cal D}^{-1}
| {\bf k}, p \rangle.
\end{equation}
\end{widetext}

As for Eq.~(\ref{fii}), the sequence of events associated to these loops may be read from 
right to left in the r.-h.-s. of (\ref{fi}): first-order rough reflection 
by mirror $j$, one-way propagation to mirror $[j+1],$
specular reflection, one-way back to $j$, second first-order rough reflection by $j.$
To restore the initial sense of propagation, the loop is closed by  one or more
specular round-trips. Before each rough reflection, arbitrary numbers of specular round-trips 
are allowed. 
Explicit evaluation yields
\begin{widetext}
\begin{equation}\label{fi_explicit}
\delta f^{(2\,\rm i)}_p({\bf k},\omega)=\frac{1}{A}\sum_{j=1}^2
\sum_{p'}
\int\frac{d^2 k'}{(2\pi)^2}
\frac{e^{-2(\kappa+\kappa')L}\,
{r}^{p}_{j+1}({\bf k})\,{r}^{p'}_{j+1}({\bf k}')\,
\langle {\bf k},p| \delta{\cal R}_j^{(1)}(\omega)| {\bf k}', p' \rangle
\langle {\bf k}',p'| \delta{\cal R}_j^{(1)}(\omega)| {\bf k}, p \rangle
}
{\left(1-r^p_{1}({\bf k},\omega)r^p_{2}({\bf k},\omega)e^{-2\kappa L}\right)^2 
(1-r^{p'}_{1}({\bf k}',\omega)r^{p'}_{2}({\bf k}',\omega)e^{-2\kappa' L})},
\end{equation}
\end{widetext}
where $\kappa'=-i\sqrt{\omega^2/c^2-k'{}^2}.$

Before replacing all these results into (\ref{Fcas}), we first write the Casimir force as the 
real part of  integrals of the loop functions~\cite{lossy_cavities}. 
Since these functions are analytical, 
Cauchy theorem allows us  to 
replace  the integral over real frequencies by an integral over the imaginary axis in the complex plane of 
frequency. As a result, $\omega$ is replaced by $\xi=-i\omega,$ and 
$\exp(-\kappa L)$ becomes a real exponential factor, with 
\[
\kappa = \sqrt{k^2+\frac{\xi^2}{c^2}} >0.
\] 
The resulting integrals turn out to be real, and  
the roughness correction is then given by 
\begin{widetext}
\begin{equation}\label{Fcas2}
\delta F_{\rm PP}(L) = 2A \int\frac{d^2 k}{(2\pi)^2}
\int_0^{\infty}\frac{d \xi}{2\pi}\hbar\kappa
\sum_p\left[ \delta f^{\rm (2\,i)}_p({\bf k},\xi)+\delta f^{\rm (2\,ii)}_p({\bf k},\xi) \right]
\end{equation}
\end{widetext}

According to 
(\ref{fii_explicit}) and (\ref{R2}), $\delta f^{\rm (2\,ii)}_p({\bf k},\xi)$
is given by an integral over ${\bf k}'$ with the integrand proportional to
$|H_j({\bf k}-{\bf k}')|^2.$ We also obtain this factor when computing $\delta f^{\rm (2\,i)}_p({\bf k},\xi)$
from 
 (\ref{R1}) and 
(\ref{fi_explicit}). 
According to (\ref{area2}), when averaged it  
yields $A\sigma_{jj}({\bf k}-{\bf k}'),$ 
due to translational symmetry on the $xy$ plane.
Hence
both loop functions are independent of $A$, yielding a 
force proportional to $A$ as expected.

The energy correction is  computed  from (\ref{Fcas2}) by a simple integration:
\[
\delta E_{\rm PP}(L)=-\int_L^{\infty} \delta F_{\rm PP}(L') \,dL'.
\]
In order to simplify the notation, 
we consider two mirrors made of the same metal, and hence with the 
same optical properties (otherwise the correction is given by 
a trivial extension). Then, after  transforming ${\bf k}-{\bf k}'$ 
 into ${\bf k}$ by a trivial change of integration variable
and
taking (\ref{area2}) into account, we find
\begin{equation}\label{main}
\delta {E}_{\rm PP} = \int\frac{d^2{\bf k}}{(2\pi)^2} \, G({\bf k})\, \sigma({\bf k}),
\end{equation}
with $\sigma({\bf k})=\sigma_{11}({\bf k})+\sigma_{22}({\bf k}).$  
As discussed in connection with Eq.~(\ref{R1}),
 the ${\bf k}$ in $G({\bf k})$ represents the field momentum  
transfer induced by a given Fourier component of the surface profile. 
 The second-order roughness response function $G({\bf k})$ is given by the following general expression
\begin{align}\label{G}
G\left( \mathbf{k}\right)  &  =  -\hbar A\int\limits_{0}^{\infty}%
\frac{d\xi}{2\pi}\int\frac{d^{2}\mathbf{k}^{\prime}}{4\pi^{2}%
}~b_{\mathbf{k',k^{\prime}-k}} \\
b_{\mathbf{k',k''}} &
= b^{\rm(i)}_{\mathbf{k',k''}}(\xi)
+ b^{\rm(ii)}_{\mathbf{k',k''}}(\xi).
\end{align}
The contribution of the first-order reflection operator is calculated from 
Eq.~(\ref{fi_explicit}) (from now on we drop 
the index $j$ indicating one of the two mirrors):
\begin{widetext}
\begin{equation}
\label{Gi}
b^{\rm(i)}_{\mathbf{k',k''}}(\xi)=\frac{1}{2}
\sum_{p'p''}\frac{
e^{-2(\kappa'+\kappa'') L}\,
r^{p'}({\bf k}',\xi)\,r^{p''}({\bf k}'',\xi)\,
 R^{(1)}_{p'p''}(\xi;{\bf k}',{\bf k}'')\,
 R^{(1)}_{p''p'}(\xi;{\bf k}'',{\bf k}')}
{(1-r^{p'}({\bf k}',\xi)^2\, e^{-2\kappa' L})\,
(1-r^{p''}({\bf k}'',\xi)^2\, e^{-2\kappa'' L})
},
\end{equation}
\end{widetext}
whereas Eq.~(\ref{fii_explicit}) leads to the 
following contribution from the second-order operator:
\begin{equation}
\label{G_ii}
b^{\rm(ii)}_{\mathbf{k',k''}}(\xi) = \sum_p
\frac{e^{-2\kappa' L}\,r^{p}({\bf k}',\xi)\,
 R^{(2)}_{p}(\xi;{\bf k}',{\bf k}'')}
{1-r^{p}({\bf k}',\xi)^2 e^{-2\kappa' L}}.
\end{equation}

The response function $G({\bf k})$ 
is entirely determined by the mirrors' non-specular coefficients 
$R^{(1)}_{pp'}$ and
$R^{(2)}_{p},$ together with the specular reflection coefficients and the 
exponential factors describing round-trip propagation inside the cavity. 
For isotropic material media, symmetry requires the response function to depend only on the modulus $k=|{\bf k}|.$
According to (\ref{main}),
this $k$ dependence describes the spectral sensitivity of the Casimir energy to roughness.
 Hence, in general the Casimir energy depends on the details of the roughness 
spectrum $\sigma({\bf k})$, and not only on the roughness variance
\[
\langle h_1^2 + h_2^2 \rangle = \int \frac{d^2{\bf k}}{(2\pi)^2}\sigma({\bf k}).
\]
 When $G(k)$ 
is known and the roughness spectrum measured experimentally, Eq. (\ref{main}) 
 allows for a precise and straightforward calculation of the roughness correction to the Casimir force without using the proximity force approximation. 

The PFA is recovered only when the 
the surface is very smooth, corresponding to a roughness 
spectrum $\sigma({\bf k})$  sharply peaked 
around ${\bf k}={\bf 0}.$  In this case, we may replace 
$G(k)$ by $G(0)$  in (\ref{main}) to find 
\begin{equation}
\delta {  E}_{\rm PP} \approx G(0)\langle h_1^2 + h_2^2 \rangle,\;\;\;\;\;\;\;\;\;\;{\rm (PFA)}
\end{equation}
in agreement with Eq.~(\ref{PFA1}) provided that the response function satisfies the limit
\begin{equation}\label{PFA2}
G(k\rightarrow 0)= \frac{E_{\rm PP}''(L)}{2},
\end{equation}
 with the Casimir energy 
in the ideal case given by \cite{Lambrecht2000}
\begin{equation}\label{plane-plane}
E_{\rm PP}(L) =  \hbar A \int\limits_{0}^{\infty}\frac{d
\xi}{2\pi}
\int\frac{d^2{\bf k}}{(2\pi)^2}\sum_{p}
\ln \left( 1-r^p({\bf k},\xi)^2e^{-2\kappa L}\right).%
\end{equation}
As a consequence of 
general properties of the rough reflection coefficients at zero momentum transfer 
(specular limit), 
we show in appendix \ref{AppB} that this limit is satisfied by any response function derived
from (\ref{G}) regardless of the model considered for the material medium.

\section{Roughness response function for the plasma model}

In this section, we present an explicit computation of the response function $G(k),$ starting from the general 
result given by Eqs.~(\ref{G})-(\ref{G_ii}), and
taking the  plasma model to describe the optical properties of the metallic mirrors. The 
dielectric function is given by
\[
\epsilon = 1 + \frac{\omega_{\rm P}^2}{\xi^2}.
\]
The plasma wavenumber, wavelength and frequency are related by 
\[
k_{\mathrm{P}}   
=\frac{2\pi}{\lambda_{\mathrm{P}}}=\frac{\omega_{\mathrm{P}}}{c}.
\]
We also define
\[
\kappa_{t}({\bf k},\xi) =\sqrt{\mathbf{k}^{2}+\epsilon\,\frac{{\xi}^{2}}{c^2}}
=\sqrt{\kappa^{2}+k_{\mathrm{P}}^{2}}
\]
representing the imaginary part of the $z$ component of the wavevector inside the metallic medium. 
The specular reflection coefficients are given by 
\begin{equation}
\label{rTE}
r^{\scriptscriptstyle\rm TE}({\bf k},\xi) = -\frac{\kappa_t-\kappa}{\kappa_t+\kappa},
\end{equation}
\begin{equation}
\label{rTM}
r^{\scriptscriptstyle\rm TM}({\bf k},\xi) = \frac{\left(1+\frac{\omega_{\rm P}^2}{\xi^2}\right)\kappa-\kappa_t}
{\left(1+\frac{\omega_{\rm P}^2}{\xi^2}\right)\kappa+\kappa_t}.
\end{equation}

In order to compute the roughness reflection coefficients,
 we follow the perturbation approach of Ref.~\cite{greffet}, 
which is based on the extinction theorem~\cite{agarwal} and the 
Rayleigh hypothesis. 
The incident, reflected and transmitted fields are related by two integral equations, 
which are solved up to second order of  $H_j({\bf k})$ 
for the reflected field in terms of the incident field. This allows us to derive
the non-specular
coefficients $R_{pp'}^{(1)}({\bf k},{\bf k}';\xi)$ and 
$R_{p}^{(2)}({\bf k},{\bf k};\xi)$ defining the relevant matrix elements of the 
first and second-order reflection operators.  

However, it turns out to be simpler to first calculate 
the coefficients $\Lambda_{pp'}^{(1)}(\xi;{\bf k},{\bf k}')$ defined as follows:
\begin{equation}
R_{pp'}^{(1)}({\bf k},{\bf k}';\xi)= \frac{r^p({\bf k},\xi)t^{p'}({\bf k}',\xi)}{t^p({\bf k},\xi)}
\Lambda_{pp'}^{(1)}(\xi;{\bf k},{\bf k}'),
\end{equation}
where $t^p({\bf k},\xi)$ are the transmission coefficients for the plane interface (see Appendix A). 

For given values of ${\bf k}$ and ${\bf k}'$, we cast the four coefficients 
$\Lambda_{pp'}^{(1)}(\xi;{\bf k},{\bf k}')$
into the $2\times 2$ matrix ${\bf \Lambda}^{(1)}({{\bf k},{\bf k}'})$ 
(with the association ${\rm TE}=1,$ ${\rm TM}=2$),  whose
nondiagonal elements  represent the coupling between TE and TM polarizations.
We find
\[
{\bf \Lambda}^{(1)}({{\bf k},{\bf k}'})={\bf \Lambda}^{(1)}_-({{\bf k},{\bf k}'}) - {\bf \Lambda}^{(1)}_+({{\bf k},{\bf k}'}),
\]
\begin{equation}
\label{Psi}
{\bf \Lambda}^{(1)}_{\pm}({{\bf k},{\bf k}'}) = \left(  \kappa_{t}\pm \kappa\right)  
{\bf B}_{t}^{-1}\left(
\begin{array}
[c]{cc}%
C & S\\
-\frac{S}{1\pm \beta\beta_{t}} & \frac{C\pm \beta\beta_{t}^{\prime}}{1\pm \beta\beta_{t}}%
\end{array}
\right)  {\bf B}_{t}^{\prime},
\end{equation}
$C={\bf k}\cdot {\bf k}'/(k\, k')$ and 
 $S=\sqrt{1-C^2}.$ 
We have also defined 
\begin{align}
\beta  &  = \frac{k}{\kappa}\qquad \beta_{t}=\frac{k}{\kappa_{t}}\nonumber\\
{\bf B}_{t}  &  =\left(
\begin{array}
[c]{cc}%
1 & 0\\
0 & \frac{c\kappa_{t}}{\sqrt{\epsilon} \xi}%
\end{array}
\right).\nonumber
\end{align}
Primed quantities are likewise defined in terms of ${\bf k}'.$  

The second-order coefficients are written in a similar way, with 
\begin{equation}
R_{p}^{(2)}({\bf k},{\bf k}';\xi)= r^p({\bf k},\xi)
\Lambda_{p}^{(2)}(\xi;{\bf k},{\bf k}').
\end{equation}
$\Lambda_{\rm\scriptscriptstyle TE}^{(2)}(\xi;{\bf k},{\bf k}')$ and 
$\Lambda_{\rm\scriptscriptstyle  TM}^{(2)}(\xi;{\bf k},{\bf k}')$
 are the diagonal elements of the matrix
\begin{equation}
{\bf \Lambda}^{(2)}({{\bf k},{\bf k}'})  =2\kappa \kappa_{t} {\bf I}
-{\bf \Lambda}^{(1)}({{\bf k},{\bf k}'}) {\bf \Lambda}^{(1)}_-({{\bf k},{\bf k}'})
\label{Delta2}
\end{equation}
with
${\bf I}$ denoting the $2 \times 2$ identity matrix. 

By replacing these results into 
(\ref{Gi}) and 
(\ref{G_ii}), we find the explicit expressions for the 
functions $b^{\rm(i)}_{\mathbf{k,k}^{\prime}}(\xi)$ and $b^{\rm(ii)}_{\mathbf{k,k}^{\prime}}(\xi):$
\begin{widetext}
\begin{align}
b^{\rm(i)}_{\mathbf{k,k}^{\prime}}(\xi)  &=& 
\frac{1}{2}\sum_{\epsilon,\epsilon'=+,-}\mu_{\epsilon}\mu_{\epsilon'}'
\Biggl[ f_{\scriptscriptstyle\rm TE}({\bf k},\xi)\,
f_{\scriptscriptstyle\rm TE}({\bf k'},\xi)\,C^{2}\left(  1+\epsilon \beta\beta_{t}\right)  \left(
1+\epsilon^{\prime}\beta^{\prime}\beta_{t}^{\prime}\right)
+f_{\scriptscriptstyle\rm TE}({\bf k},\xi)\, f_{\scriptscriptstyle\rm TM}({\bf k'},\xi)\,S^{2}\left(  1+\epsilon \beta\beta_{t}\right)
\label{g1}\\
&   & 
+f_{\scriptscriptstyle\rm TM}({\bf k},\xi)\, f_{\scriptscriptstyle\rm TE}({\bf k'},\xi)\,S^{2}
\left(  1+\epsilon^{\prime}\beta^{\prime}\beta_{t}^{\prime
}\right)   + f_{\scriptscriptstyle\rm TM}({\bf k},\xi)\, f_{\scriptscriptstyle\rm TM}({\bf k'},\xi)\,
\left(  C+\epsilon \beta\beta_{t}^{\prime}\right)  \left(
C+\epsilon^{\prime}\beta^{\prime}\beta_{t}\right)\Biggr],\nonumber
\end{align}
\begin{align}\label{g2}
b^{\rm(ii)}_{\mathbf{k,k}^{\prime}}(\xi)  = & 
2\,\kappa_{t}\,\kappa\,\left(  f_{\scriptscriptstyle\rm TE}({\bf k},\xi)
+f_{\scriptscriptstyle\rm TM}({\bf k},\xi)\right) 
 +\sum_{\epsilon=+,-}\mu_{\epsilon}\mu_{-}'   
\Biggl[
  f_{\scriptscriptstyle\rm TE}({\bf k},\xi)\,\left(1-C^{2} \beta' \beta_t'\right) \left(  1+\epsilon \beta\beta_{t}\right)   
   \\
 &  +f_{\scriptscriptstyle\rm TM}({\bf k},\xi)\,S^{2}\left(
1-\beta^{\prime}\beta_{t}^{\prime}\right) +f_{\scriptscriptstyle\rm TM}({\bf k},\xi)\,
\left(  C+\epsilon \beta\beta_{t}^{\prime}\right)  \left(  C-\beta^{\prime}\beta_{t}\right)\Biggr],\nonumber
\end{align}
\end{widetext}
with
\[
\mu_{\pm}  =\frac{\kappa\pm\kappa_{t}}{1\pm \beta\beta_{t}}.
\]

These expressions can now be applied to the numerical computation of
the response function for arbitrary values of 
$L$ and $\lambda_{\rm P}$. In Fig.~3, we plot $G/E_{\rm PP}$ as a function of $k$ for several different values of the 
distance $L,$ and for $\lambda_{\rm P}=136 {\rm nm},$ which corresponds to gold covered mirrors.

\begin{figure}[ptb]
\centerline{\psfig{figure=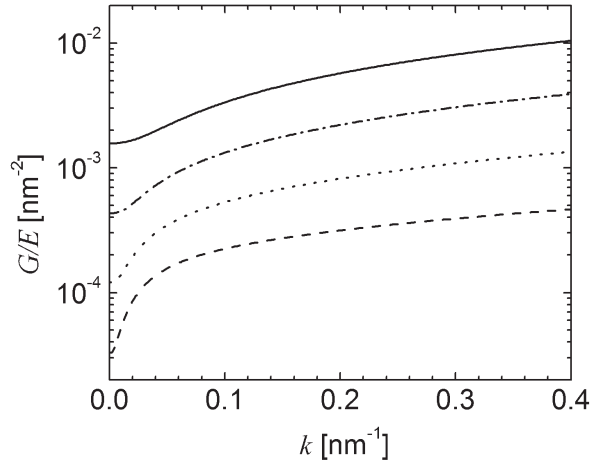,width=9cm}}\caption{Variation of $G/E_{\rm PP}$
versus $k$ 
for the distances $L=50 {\rm nm}$ (solid line), 
$L=100 {\rm nm}$ (dashed-dotted line), $L=200 {\rm nm}$ (dotted line), 
and 
$L=400 {\rm nm}$ (dashed line). We take $\lambda_{\rm P}=136 {\rm nm}.$ }
\end{figure}

According to (\ref{main}), this ratio provides the relative correction of the Casimir energy in the 
plane-plane configuration when integrated over the roughness spectrum $\sigma({\bf k}).$
Moreover, from (\ref{PFAsphere}) it also provides the relative force correction 
$\Delta$
for the plane-sphere 
geometry when the sphere radius is sufficiently large. Fig.~3 indicates that the relative correction is larger for shorter 
distances. 

The behavior of $G(k)$ as $k\rightarrow 0$ is related to the PFA, and was already discussed 
in connection with Eq.~(\ref{PFA2}). In addition to the 
verification of the general $G(k)$ given by (\ref{G})-(\ref{G_ii}) (see Appendix B),
we have also verified independently that the explicit 
result derived from (\ref{g1}) and (\ref{g2}) 
also agrees with Eq.~(\ref{PFA2}). 

The fact that $G$ increases as 
$k$ grows from zero, as displayed in Fig.~3, implies that the PFA underestimates the roughness correction.
In order to quantify the departure from the PFA description, we define the sensitivity function
\begin{equation} \label{definition_rho}
\rho(k) = \frac{G(k)}{G(0)}.
\end{equation}
In Fig.~4, we plot $\rho$ as function of $k$ for the same values of distance and $\lambda_{\rm P}$ 
employed in Fig.~3. The PFA amounts to replace $\rho(k)$ by unity for all values of 
$k$ contained in the roughness spectrum $\sigma(k).$ Clearly, this approximation is better 
for shorter distances, and smaller values of $k$ (corresponding to longer roughness wavelengths), 
as expected. For instance, the inlet shows that the PFA is a good approximation for $L=50 {\rm nm}$ and 
$k<0.04 {\rm nm}^{-1}.$ On the other hand, for $L=200 {\rm nm}$ and $k=0.02 {\rm nm}^{-1}$ 
(roughness wavelength $2\pi/k \simeq 300 {\rm nm}$) we find 
$\rho \simeq 1.6,$ corresponding to a roughness correction $60\%$ larger than the PFA result. 

\begin{figure}[ptb]
\centerline{\psfig{figure=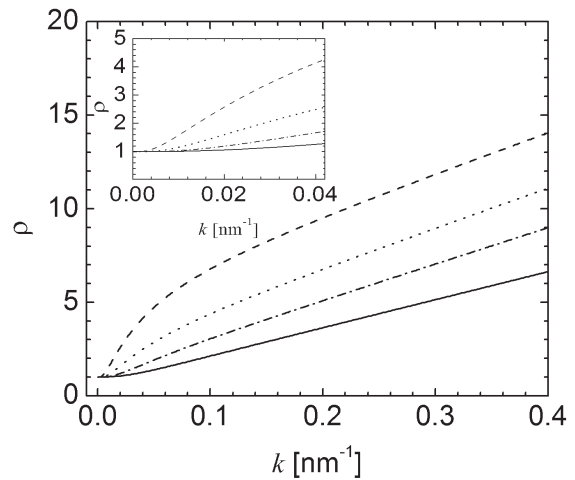,width=9cm}}\caption{Variation of $\rho$
versus $k$ for several values of $L$ (same conventions as on Fig.~3).}
\end{figure}

Fig.~4 also indicates that $\rho(k)$ grows linearly for large values of $k.$ In the next section, 
we show that this is a general result, valid for arbitrary values of $L$ and $\lambda_{\rm P}.$

\section{High-$k$ limit}

When the momentum transfer $k$ is much larger than $1/L,$
the function $b^{\rm(i)}_{\mathbf{k',k'-k}}(\xi),$
representing the 
contribution of the first-order reflection operator,
is negligible, because
it is proportional to the 
exponentially small propagation factor $\exp(-\sqrt{({\bf k'}-{\bf k})^2+\xi^2} L)\approx \exp(-k L)$
appearing in Eq.~(\ref{Gi}). This general property can be understood from the discussion 
of Sec~III: the loop function 
$\delta f_p^{(2\rm i)}({\bf k}',\xi)$
contains two rough reflections separated by a intracavity round-trip propagation with the modified momentum  
${\bf k'}-{\bf k}.$  

On the other hand, the loop function 
$\delta f_p^{(2\rm ii)}({\bf k},\xi)$
 involves a single second-order rough reflection, 
which must conserve momentum so as to allow for a closed loop. Thus, 
$b^{\rm(ii)}_{\mathbf{k',k'-k}}$ does not involve propagation with the modified momentum
${\bf k'-k}$ and is the dominant term in the high-${\bf k}$ limit.  
We calculate 
$b^{\rm(ii)}_{\mathbf{k',k'-k}}$ from 
(\ref{g2})
by taking ${\bf k'-k}\approx -{\bf k}.$ We  also assume that 
$k\gg k_{\rm P};$ the opposite case will be discussed in Sec.~VI. 
We find
\begin{equation}
\rho(k)= \alpha\, k \quad {\rm for}\quad k^{-1} \ll \lambda_{\rm P}, L.
\end{equation}
The dimensionless parameter $\alpha/L$ depends on
$K_{\rm P}=k_{\rm P} L = 2\pi L/\lambda_{\rm P}$ only, and is given by  
\begin{widetext}
\begin{equation}\label{high-k}
\alpha=\frac{\hbar c A}{(2\pi)^2 L^4 G(0)}
\int_0^{\infty} d\gamma \gamma \int_0^\gamma d\Omega
\frac{K_{\rm P}^2}{2\Omega^2+K_{\rm P}^2}\left[\gamma f_{\scriptscriptstyle\rm TE}({\bf k},\xi)+
  \frac{2(\gamma^2-\Omega^2)^2-\gamma_t^2(2\gamma^2-3\Omega^2)}
 {(\gamma \gamma_t)^2-(\gamma^2-\Omega^2)^2}
\gamma f_{\scriptscriptstyle\rm TM}({\bf k},\xi)\right].
\end{equation}
\end{widetext}
We have introduced the dimensionless integration variables
$\gamma=\kappa L,$ $\gamma_t= \kappa_t L,$ and $\Omega = \xi L/c.$ 
$f_{\scriptscriptstyle\rm TE}({\bf k},\xi)$ and 
$f_{\scriptscriptstyle\rm TM}({\bf k},\xi)$ are calculated 
from (\ref{fp}) as
functions of $\gamma$ and $\Omega.$ 

We plot the coefficient $\alpha$ as a function of $L$ in Fig.~5,
with the plasma wavelength of gold $\lambda_{\rm P}=136$nm as in the previous 
numerical examples.
At the limit of short distances, we recover from (\ref{high-k}) 
our previous result \cite{EPL} 
\begin{equation} \label{kgde-plasmon}
\alpha = 0.4492 L \quad {\rm for} \quad k^{-1}\ll L\ll \lambda_{\rm P}.
\end{equation}
This corresponds to the high-$k$ limit of the plasmon (non-retarded) regime, which we shall discuss
further in Sec.~VII. This limit is indicated by the dotted line in Fig.~5. 

As shown by Fig.~5, the angular coefficient $\alpha$ saturates at the limit of large distances.
This corresponds to the limit $k^{-1}\ll \lambda_{\rm P}\ll L,$
which may be obtained analytically from Eq.~(\ref{high-k})
by expanding its r.-h.-s. in powers of $\lambda_{\rm P}.$
The integrand  vanishes to order $\lambda_{\rm P}^{-1},$ whereas the 
the zero-th order term yields
\begin{align}
\alpha^{(0)} = & -\frac{\hbar c A}{(2\pi)^2 L^4 G(0)}\int_0^{\infty} d{\gamma}\frac{1}{e^{2{\gamma}}-1}\\
 &\times\int_0^{\gamma} d{\Omega}({\gamma}^2-3{\Omega}^2)=0,\nonumber
\end{align}
so that the dominant term is of the order of $\lambda_{\rm P}.$
We also need to calculate $G(0)$ in the limit $L\gg \lambda_{\rm P}.$ 
As expected, we find $G(0)=E^{\rm pr}_{\rm PP}{}''(L)/2=-\pi^2\hbar c A/(120 L^5),$
where $E_{\rm PP}^{\rm pr}$ is 
the Casimir energy for perfectly 
reflecting mirrors.  We then find  
\begin{align}
\alpha = & \frac{60}{\pi^5} \lambda_{\rm P} \int_0^{\infty} d{\gamma}\frac{e^{2{\gamma}}}
{{\gamma}(e^{2{\gamma}}-1)^2}\label{an1}\\
 &
 \times\int_0^{\gamma} d{\Omega}({\gamma}^4-2{\gamma}^2{\Omega}^2+3{\Omega}^4)\nonumber
\end{align}
giving
\begin{equation}\label{limit_alpha}
\alpha=\frac{7}{15\pi} \lambda_{\rm P} \quad{\rm for} \quad k^{-1} \ll \lambda_{\rm P} \ll L,
\end{equation}
which is in agreement with the saturation value shown in Fig.~5.
This result remarkably  differs from the long distance behavior reported in Ref.~\cite{EPL}, 
which corresponds to the perfectly-reflecting limit. 
Note that the high-$k$ expression (\ref{high-k}) holds when the roughness length scale 
$1/k$ is much smaller than both $\lambda_{\rm P}$ and $L$. In this regime, and
as a consequence of the momentum transfer induced by the roughness effect, the modified field momentum 
has a magnitude $|{\bf k' - k}|$ much larger than $k_{\rm P}.$ Therefore, it is poorly reflected by the mirrors, 
even though the initial momentum satisfies $k'\stackrel{<}{\sim} 1/L\ll k_{\rm P}.$
In order to obtain the perfectly-reflecting limit, one must assume that 
  $\lambda_{\rm P}$ rather than $1/k$ is the shortest length scale, as discussed in the next
section.

\begin{figure}[ptb]
\centerline{\psfig{figure=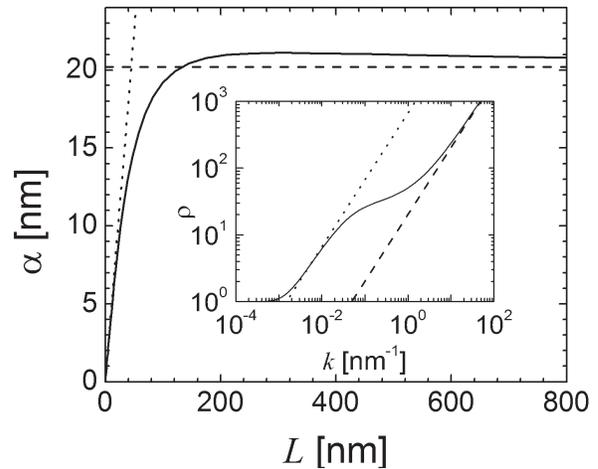,width=9cm}}\caption{Variation of the 
angular coefficient 
$\alpha$ versus $L$ for $\lambda_{\rm P}=136$nm. The analytical result for 
$k^{-1} \ll L \ll \lambda_{\rm P}$ is shown as the dotted line and for
$k^{-1} \ll \lambda_{\rm P} \ll L$ as the dashed line.   
A comparison between this second result (dashed straight line) 
and the exact $\rho(k)$ (solid line) is shown in the inlet for $L=2 \mu$m. 
The analytical result $\rho=L k/3$ predicted by the model of perfect reflectors
(dotted line) is valid only in the intermediate range $ \lambda_{\rm P}\ll k^{-1} \ll L$.}
\end{figure}

\section{Perfectly-reflecting limit}

In this section we assume that $\lambda_{\rm P}$ is much smaller than both 
the separation $L$ and the roughness wavelength  $1/k.$ Then, we expand 
$b_{\bf k', k''}$ in powers of $\lambda_{\rm P}.$ For $b^{\rm(ii)}_{\bf k', k''},$
representing the
contribution of the second-order 
reflection operator and given by (\ref{g2}), the dominant term is of the order of 
$1/\lambda_{\rm P}:$
\begin{equation}\label{g2-1}
b^{\rm(ii)}_{\bf k', k'-k}=4\pi\frac{f^{\rm pr}({\bf k'})}{\kappa'\,\lambda_{\rm P}}\mathbf{k'\cdot k}+
O\left( \lambda_{\rm P}^{0}\right),
\end{equation}
where  $f^{\rm pr}({\bf k'})$ represents  the loop function for perfect 
reflectors [it is the same for both polarizations, with $r^{\scriptscriptstyle\rm TE}=
-r^{\scriptscriptstyle\rm TM}=-1,$ according to (\ref{rTE}) and (\ref{rTM})].
On the other hand, $b^{\rm(i)}_{\bf k', k' -k}$
vanishes up to order $1/\lambda_{\rm P}.$
It follows that $G\left(\mathbf{k}\right)$ vanishes at this order, because when taking the 
integral of the r.-h.-s. of (\ref{g2-1}) over all values of momentum ${\bf k'}$ in (\ref{G}), opposite values of  
$\mathbf{k'}$ compensate each other.

Both $b^{\rm(i)}_{\bf k', k''}$
and $b^{\rm(ii)}_{\bf k', k''}$
contribute up to order $\lambda_{\rm P}^0.$ 
It is useful to replace $b_{\bf k', k''}$ by the symmetrized 
form 
\[
{\hat b}_{\bf k', k''}=(b_{\bf k', k''}+b_{\bf k'', k'})/2.
\]
This procedure does not change the response function because 
$G({\bf k})=G(-{\bf k})=G(k).$
Taking $d^2{\bf k'}=dk' k' d\phi'$ in (\ref{G}), we derive
\begin{equation}\label{perfeito}
G({\bf k})   = -\frac{\hbar A}{8\pi^3}\int\limits_{0}^{\infty}d\xi\int_{0}^{\infty}
dk' \,k' \int_{0}^{2\pi}d\phi'
~{{\hat b}_{\mathbf{k', k'-k}}}.
\end{equation}
${\hat b}_{\bf k', k''}$ is obtained from the term of order $\lambda_{\rm P}^0$
in (\ref{g1}) and (\ref{g2}): 
\begin{align}\label{hatg}
{{\hat b}_{\mathbf{k', k''}}}  &  =\frac{e^{-2\kappa' L}+
e^{-2\kappa'' L}}{\left(  1-e^{-2\kappa' L}\right)  \left(  1-e^{-2\kappa''%
L}\right)  }\\
&  \times\frac{\left(  \kappa'\kappa''\right)  ^{2}+\left(  
{\xi}^{2}/c^2+\mathbf{k'\cdot k''}\right)  ^{2}}{\kappa'\kappa''}.\nonumber
\end{align}
We change the variables of integration from $(\xi,k')$ to 
$(\kappa',\kappa'').$ The integral over $\phi'$ yields
\begin{equation}\label{angular}
\int_0^{\phi_m} |J|\; k(\kappa',\kappa'',\phi')d\phi' = \frac{\pi}{2}\frac{\kappa'\kappa''}{k},
\end{equation}
where $\phi_m={\rm arcsin}[(\kappa'{}^2-\kappa''{}^2+k^2)/(2\kappa' k)]$
and $|J|$ is the Jacobian corresponding to the transformation.
 
From (\ref{perfeito})--(\ref{angular}) we derive 
\begin{align}\label{Gperfect}
G(k)= &-\frac{\hbar c A}{8\pi^2}\,\frac{1}{L^5\,q}\int_0^{\infty}\frac{d\gamma e^{-2\gamma}}{1-e^{-2\gamma}}
\int_{|\gamma-q|}^{\gamma+q}d\gamma'\\
\times  & 
\frac{(\gamma\gamma')^2+\frac{1}{4}(\gamma^2+\gamma'{}^2-q^2)^2}
{1-e^{-2\gamma'}} \quad {\rm for} \; \lambda_{\rm P} \to 0.\nonumber
\end{align}
$\gamma$ has the same meaning already discussed in connection with (\ref{high-k}) while
$\gamma'$ corresponds to the diffracted wave. From (\ref{Gperfect}) 
we verify that $G(0)=E^{\rm pr}_{\rm PP}{}''/2$ as expected.   
 For arbitrary values of $q=kL,$ numerical integration of 
(\ref{Gperfect}) agrees with the results of Emig {\it et al.}~\cite{Emig}
for a perfectly-reflecting mirror corrugated along a fixed direction in the $xy$ plane. 
By taking the 
assumption of perfect reflectivity from the start, we may derive this result 
for a general deformation directly from 
our general expressions (\ref{G})-(\ref{G_ii}), as discussed in Appendix C.

In order to discuss the regime $\lambda_{\rm P}\ll 1/k \ll L$, we now take the high-$k$ limit 
of the right-hand side of (\ref{Gperfect}).
Due to the presence of the exponential factor $\exp(-2\gamma)$, 
the dominant contribution comes from the corner 
$\gamma\stackrel{\scriptstyle <}{\scriptstyle\sim}1,$ $\gamma'\sim q$  
of the rectangle associated to the integration region.  
We may thus neglect $\exp(-2\gamma')$ and 
recover the long distance limit of \cite{EPL}: 
\begin{eqnarray}
\label{perfect}
G(k)&=&-\frac{2}{3\pi^2} \frac{\hbar c A q}{L^5} \int_0^{\infty} d\gamma \frac{\gamma^3
\,e^{-2\gamma}}{1-e^{-2\gamma}}=-\frac{\pi^2}{360}\hbar A\frac{q}{L^4}, \nonumber \\
\rho&=&\frac{1}{3}L\,k \quad{\rm for} \quad \lambda_{\rm P} \ll k^{-1} \ll L. 
\end{eqnarray}

In summary, the long-distance behavior is given by (\ref{limit_alpha}) 
when $1/k\ll \lambda_{\rm P}\ll L,$ and by (\ref{perfect}) when $\lambda_{\rm P}\ll 1/k\ll L$.
The cross-over between these two regimes is shown in the inlet of Fig.~5, where
we plot $\rho$ as a function of $k$ for $L=2 \mu$m. 
The finite conductivity of the metals clearly reduces the roughness correction 
for very large values of $k L,$ due to the saturation effect discussed in Sec.~V. 
In Fig.~6, we  plot  
$\rho$ as function of the plasma 
wavelength
$\lambda_{\rm P}$ for $k=0.02 {\rm nm}^{-1}$ (roughness
wavelength $2\pi/k\approx 300 {\rm nm}$), with $L= 2\mu{\rm m}$ (solid line) and
$L=600{\rm nm}$ (dashed line), in order to analyze in detail the effect of finite conductivity.
The value $\lambda_{\rm P}=136 {\rm nm}$ corresponding
to gold covered mirrors is highlighted by a vertical line.  
The perfeclty-reflecting limit corresponds to the small values of $\lambda_{\rm P}$ shown in the left-hand side of the figure.  The correction decreases as $\lambda_{\rm P}$ approaches 
$2\pi/k$ because of the saturation effect, and then it increases again as $\lambda_{\rm P}$ approaches and goes beyond 
the  separation distance $L.$ 
For $\lambda_{\rm P}\gg L,$ we find the limit predicted by the plasmon model, to be discussed in the next section, 
which is larger than the perfeclty-reflecting limit~\cite{EPL}.
When $k L \stackrel{\scriptscriptstyle <}{\sim} 1,$ 
the  correction is always larger than the perfectly-reflecting limit, as illustrated 
in the inlet of Fig.~6, where we take $k=0.02 {\rm nm}^{-1}$
and $L= 100{\rm nm}.$  In this case, the perfectly-reflecting limit is $11\%$ larger than the PFA result,
 whereas at $\lambda_{\rm P}=136 {\rm nm}$ we find a correction $19\%$ larger than the PFA.

\begin{figure}[ptb]
\centerline{\psfig{figure=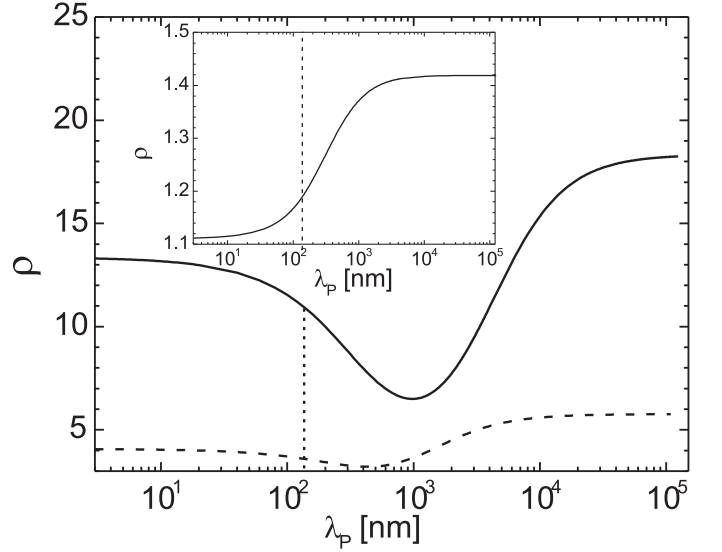,width=9cm}}\caption{Variation of $\rho$
versus $\lambda_{\rm P}$ for $k=0.02 {\rm nm}$ and the distances 
$L=2\mu{\rm m}$ (solid line), $L=600{\rm nm}$ (dashed line) and $L=100{\rm nm}$ (inlet, solid line).
The values at $\lambda_{\rm P}=136 {\rm nm}$ are indicated by vertical lines.}
\end{figure}

\section{Plasmon limit}

In the short distance regime, $L \ll \lambda_{\rm P},$ the Casimir
energy is associated to surface plasmons. As discussed in the previous sections, 
when integrating  $b_{\bf k', k'-k}$ in (\ref{G}), the dominant contributions come from 
values of ${\bf k'}$ and $\xi$ such that $\kappa'\stackrel{\scriptstyle <}{\scriptstyle\sim} 1/L.$
Then, the short-distance limit of (\ref{rTE}) yields $r^{\rm\scriptscriptstyle TE} = O(L/\lambda_{\rm P})^2,$ so that
the contribution of TE polarization is negligible. From (\ref{rTM}), $r^{\rm\scriptscriptstyle TM}$
 is also negligible 
except for the values 
\begin{equation}\label{cond-plasmon}
\xi\stackrel{\scriptstyle <}{\scriptstyle\sim} \omega_{\rm P} \ll c/L,
\end{equation}
for which $\kappa' \approx k',$ and 
\begin{equation}\label{rm}
r^{\rm\scriptscriptstyle TM} \approx \frac{\omega_{\rm P}^2}{2\xi^2+ \omega_{\rm P}^2}.
\end{equation} 
 In other words, the dominant contribution is associated to low-frequency surface waves, the 
reflection coefficient having poles at the surface plasmon  resonance $\xi =  - i \omega_{\rm P}/\sqrt{2}.$ 
From (\ref{cond-plasmon}), we also conclude that the retardation time $L/c$ for propagation 
between the mirrors is negligible in the time scale associated to the relevant field frequencies. 

We calculate the roughness correction by taking the appropriate limits in (\ref{g1}) and 
(\ref{g2}). Since $r^{\rm\scriptscriptstyle TM}$ only depends on $\xi,$ the diffracted wave 
sees the same 
reflection coefficient, which we denote as $r$ for simplicity. We find 
\begin{widetext}
\begin{equation}
\label{g1plasmon}
b^{\rm(i)}_{{\bf k}',{\bf k}''} = \frac{k'k''\,r^4e^{-2k'L}e^{-2k''L}}{2(1- r^2 e^{-2k'L})(1- r^2 e^{-2k''L})}
\left[(C+1)^2+2r(1-C^2)+r^2(1-C)^2\right],
\end{equation}
\begin{equation}
b^{\rm(ii)}_{{\bf k}',{\bf k}''} =  
\frac{2k'{}^2\,r^2 e^{-2k'L}}{1- r^2 e^{-2k'L}}+ \frac{k'k''\,r^3 e^{-2k'L}}{1- r^2 e^{-2k'L}}
\left[r(1-C)^2+1-C^2\right].\label{g2plasmon}
\end{equation}
\end{widetext}
These equations do not agree with the results of Ref.~\cite{Maradudin}, which in their turn are 
not consistent with the PFA. Therefore, it is important to  check our results 
against the PFA by taking 
$k''=k' = k$ and $C=1$ in (\ref{g1plasmon}) and (\ref{g2plasmon}):
\begin{equation}
b_{{\bf k},{\bf k}}=\frac{2k^2 r^2 e^{-2kL}}{(1- r^2 e^{-2kL})^2}.
\end{equation}

When replacing this result into
(\ref{G}), and
taking (\ref{rm}) into account (with $\zeta=\xi/\omega_{\rm P},$ and  $\gamma=\kappa L=kL$), we obtain
\[
G(0) = -\frac{\hbar A}{2\pi^2} \frac{\omega_{\rm P}}{L^4}\int_0^{\infty}d\zeta\int_0^{\infty}d\gamma 
\frac{\gamma^3 (2\zeta^2+1) e^{2\gamma}}{\left[(2\zeta^2+1)^2 e^{2\gamma} -1\right]^2}
\]
Integrating this expression by parts, one shows that $G(0)$ 
satisfies (\ref{PFA2}),
with  
$E_{\rm PP}$ representing the short-distance limit of the Casimir energy as  given by~\cite{VanKampen}:
\[
E_{\rm PP}(L) = -\frac{\hbar A}{4\pi^2} \frac{\omega_{\rm P}}{L^2}\int_0^{\infty}d\zeta\int_0^{\infty}d\gamma 
\frac{\gamma^2}{(2\zeta^2+1)^2 e^{2\gamma} -1}.
\]
This confirms consistency with the PFA, which is in line with the more general discussion presented 
in Appendix B. 

In addition to the limit $k\rightarrow 0,$ it is also interesting to analyze the case
$k\gg 1/L$ from (\ref{g1plasmon}) and (\ref{g2plasmon}), allowing for the evaluation of the limit 
$1/k\ll L \ll \lambda_{\rm P}.$ This provide as additional check, by comparison with the results of 
Sec.~V.  
 In this limit, $b^{\rm(i)}_{\mathbf{k',k'-k}}$ is exponentially small, and the dominant 
contribution for $b^{\rm(ii)}_{\mathbf{k',k'-k}}$ comes form the second term 
in (\ref{g2plasmon}),  which is proportional to
$k:$
\begin{equation}
b_{{\bf k}',{\bf k}'-{\bf k}}\approx b_{{\bf k}',-{\bf k}}\approx 
\frac{k'r^3e^{-2k'L}}{1- r^2 e^{-2k'L}}\left[r(1-C)^2+1-C^2\right]k.
\end{equation}
We also take $C\approx -{\bf k'}\cdot {\bf k}/(kk')=-\cos\phi,$ and integrate over
$\phi$ to derive from  (\ref{G})
\begin{align}\label{plasmon-high-q}
G(k)  & =  -\frac{\hbar A}{8\pi^2}\frac{\omega_{\rm P} }{L^3} k \\
  & \times\int_0^{\infty}d\zeta\int_0^{\infty}d{\gamma}
\frac{{\gamma}^2 (2\zeta^2+4)}{(2\zeta^2+1)^2\left[(2\zeta^2+1)^2 e^{2{\gamma}} -1\right]},
\nonumber
\end{align}
yielding 
$
\rho(k) = 0.4492 L\, k,
$
in agreement with 
Eq.~(\ref{kgde-plasmon})
and
Ref.~\cite{EPL}.

\section{Concluding remarks}

We have calculated the second--order response function $G(k)$ for arbitrary values of the plasma
wavelength $\lambda_P$ and the distance $L.$ This allows for a reliable computation of the 
roughness correction, once the roughness spectrum $\sigma({\bf k})$ characterizing the 
metallic surfaces is experimentally determined. In order to gain further insight into the  
 the roughness correction itself, we consider the 
particularly simple example
of a Gaussian spectrum~\cite{Maradudin}: 
\begin{equation}\label{Gaussian}
\sigma\lbrack\mathbf{k}]=\pi a^{2}\ell_{C}^{2}\exp\left(
-\frac{\mathbf{k}^{2}\ell_{C}^{2}}{4}\right).  
\end{equation}
The roughness variance values $a^2,$ and $\ell_C$ represents the correlation length. 

According to Eq.~(\ref{main}), the relative force correction $\Delta$ is obtained by 
integrating the normalized response function $G(k)/E_{\rm PP}$
(see Fig.~3 for some numerical examples) over the Gaussian spectrum 
given by (\ref{Gaussian}). 
In Ref.~\cite{letter2}, we have discussed some simple analytical expressions  
in the limiting cases $\lambda_P\ll \ell_C$ and $\lambda_P\gg \ell_C,$
which can be easily derived from the results of Secs.~V--VII.
However, the experimental parameters are likely to be such that neither of the two limits
holds. Hence, we must rely on the numerical calculation of $G(k)$ to compute the correction. 
In Fig.~7, we compare the exact results for $\Delta/a^2$ 
(for two different values of $\ell_C$) with the PFA formula
 $\Delta/a^2=E_{\rm PP}''(L)/[2E_{\rm PP}(L)],$
taking as before $\lambda_P = 136 {\rm nm}.$ 
 At $L=100 {\rm nm},$ the exact value
is $57\%$  and $7\%$ larger than the PFA result
 for $\ell_C=50 {\rm nm}$ and $\ell_C=150 {\rm nm},$ respectively. 
In agreement with the discussion of Sec.~IV, the PFA is better for shorter distances and 
longer correlation lengths. For instance, as suggested by Fig.~7, it provides accurate
results if $L<100 {\rm nm}$ and $\ell_C > 150 {\rm nm}.$ Note, however, that the validity of the PFA 
can be addressed in a reliable way only from the analysis of the experimentally measured 
roughness spectrum. In the Gaussian model 
discussed in this section, the contribution of high values of $k$ are exponentially small, but in the real case
the decay of $\sigma(k)$ might be smoother. Thus, this approximation might be worse than discussed here.     

For large values of $L,$ no simple analytical result is available when 
$\lambda_P \sim \ell_C.$ In the inlet of Fig.~7, we show that the perfectly-reflecting 
result
 $\Delta/a^2= 2\sqrt{\pi}/(\ell_C L)$ valid for $\lambda_P\ll\ell_C\ll L$ 
overestimates the correction for $\ell_C=50 {\rm nm}$ by more
than $50\%.$ This is a consequence of 
the saturation effect discussed in Sec.~V: diffraction by roughness Fourier components at $k > \lambda_P^{-1}$
give rise to waves which are poorly reflected by the mirrors. Thus, when the roughness 
spectrum contains very high values of $k,$ the correction is  reduced with 
respect to the result of the perfectly-reflecting model. The reduction factor
decreases 
as $k\lambda_P\rightarrow \infty$
down to the limit $(7/5\pi)\lambda_P/L\ll 1.$
 
\begin{figure}[ptb]
\centerline{\psfig{figure=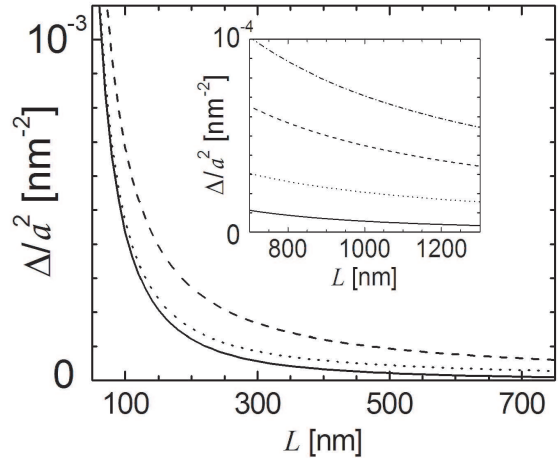,width=9cm}}
\caption{Roughness correction for the Gaussian spectrum.
We plot the relative force correction for the plane-sphere setup over the squared 
amplitude of roughness,
 $\Delta/a^2$,
versus $L$ 
for  $\ell_C=50 {\rm nm}$ (dashed line) and
$\ell_C=150 {\rm nm}$ (dotted line). The solid line represents the 
PFA result. In the inlet, we also plot   
the long-distance perfectly-reflecting limit (dot-dash). 
 We take $\lambda_{\rm P}=136 {\rm nm}.$}
\end{figure}

In conclusion, we presented 
a perturbative method for the calculation of the Casimir energy 
between rough mirrors, up
to second order 
in the amplitude of the deformation. It relies on the manipulation of 
reflection operators, taking into account diffraction and 
the coupling between different field polarizations. 
We applied the method to compute the 
roughness correction  in the framework of the plasma model,
 for arbitrary values of the plasma and roughness wavelengths
and the mirror separation $L.$ 
Analytical results for different limiting cases were discussed. 
In particular, the PFA regime follows from our formalism in the limit
of very smooth surface profiles. 
By comparison with our numerical results, we were able to analyze the accuracy
of the PFA in the problem of roughness. 
For a given roughness spectrum, 
our theory provides reliable numerical results for the roughness 
correction, and allows to check the 
validity of the PFA approach in a given experiment.
More realistic models for the metallic mirrors \cite{Lambrecht2000}
can also be considered by applying  the formal results presented here.

We thank Cyriaque Genet and Marc-Thierry Jaekel for discussions.
PAMN thanks Instituto do Mil\^enio de 
Informa\c c\~ao Qu\^antica and CNPq for partial 
financial support.

\appendix

\section{Optical network theory and the Casimir force} 

In this appendix, 
we compute the  spectral density 
in the intracavity region by generalizing the optical network formalism 
of Ref.~\cite{lossy_cavities} to the case of rough surfaces. 
This allows us to compute the Casimir force with the help of the Maxwell stress tensor.

The basic idea is to derive the relation between the intracavity field and the incoming 
outside 
field, whose fluctuations are known. 
The outside field propagating from the region $z<0$ 
(see Fig.~1)
is written as [with ${\bf r}=(x,y)$, 
$\omega = c\sqrt{k^2+k_z^2}$ and
 $\epsilon_0$ denoting the vacuum permittivity] 
\begin{align}
\label{E_in}
{\bf E}_{\rm L}^{\rm in}({\bf r},z,t) = & \sum_{p={\scriptscriptstyle \rm TE, TM}}
\int\frac{d^2{\bf k}}{4\pi^2}\int_0^{\infty} \frac{dk_z}{2\pi}
\sqrt{\frac{\hbar\omega}{2\epsilon_0}}e_{\rm L}^{\rm in}{}^p({\bf K}) \\
 & \times \exp[i({\bf k}\cdot{\bf r}+k_z z-\omega t)] {\hat \epsilon}^{p} \,+\, {\rm H. c.}. \nonumber
\end{align}
The Fourier components
 $e_{\rm L}^{\rm in}{}^p({\bf K})$ satisfy the commutation relations of freely-propagating fields:
\begin{equation}\label{commutation}
[e_{\rm L}^{\rm in}{}^p({\bf K}),e_{\rm L}^{\rm in}{}^{p'}({\bf K'}){}^{\dagger}]=
 (2\pi)^3\delta^{(2)}({\bf k - k'})\delta(k_z-k_z')\delta_{p,p'}.
\end{equation} 
The free-space fields propagating along the negative $z$-direction are written in a similar way, except
for the replacement
 $\exp(i k_z z)\rightarrow \exp(-i k_z z).$
We use the same ${\bf K}={\bf k}+k_z {\hat z}$ to label them as well, and our notation is such that 
$k_z>0$ in all cases.

We define scattering and transfer operators for the two rough mirrors and 
for the empty--space propagation between them.  The cavity is taken as a 
composed network, and the corresponding 
transfer operator is simply the product of the transfer operators for the elementary components.

\begin{figure}[ptb]
\centerline{\psfig{figure=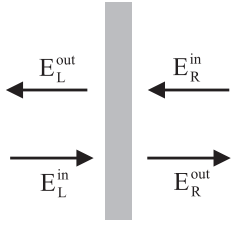,width=3cm}}\caption{Input and output fields.}
\end{figure}

The scattering operator characterizing  a given  element of the network is defined in the following way.
As shown in Fig.~8, the input field contains components propagating from the left and righthanded sides, and with
polarizations TE and TM. We arrange these components into a column vector:
\[
{\bf e}^{\rm in}({\bf K}) \equiv \left(
\begin{array}{c}
e^{\rm in}_{\rm L}{}^{\rm\scriptscriptstyle TE}\\
e^{\rm in}_{\rm R}{}^{\rm\scriptscriptstyle TE}\\
e^{\rm in}_{\rm L}{}^{\rm\scriptscriptstyle TM}\\
e^{\rm in}_{\rm R}{}^{\rm\scriptscriptstyle TM}
\end{array}
\right)_{\bf K}
\]

We employ a similar notation for the output field:
\[
{\bf e}^{\rm out}({\bf K}) =
\left(
\begin{array}{c}
e^{\rm out}_{\rm R}{}^{\rm\scriptscriptstyle TE}\\
e^{\rm out}_{\rm L}{}^{\rm\scriptscriptstyle TE}\\
e^{\rm out}_{\rm R}{}^{\rm\scriptscriptstyle TM}\\
e^{\rm out}_{\rm L}{}^{\rm\scriptscriptstyle TM}
\end{array}
\right)_{\bf K}
\]

The scattering operator provides the
input-output relation 
\begin{equation}
{\bf e}^{\rm out}({\bf K}) =
\int_{(k'_z>0)} \frac{d^3K'}{(2\pi)^3} {\bf S}({\bf K},{\bf K}'){\bf e}^{\rm in}({\bf K}') 
\end{equation}
where the integral  over $k'_z$  runs from $0$ to $\infty.$ 

Whereas Ref.~\cite{lossy_cavities}
allows for lossy mirrors, here we
assume that there is no dissipation in our network. Hence we only consider unitary scattering operators:
$ {\bf S}\cdot{\bf S}^{\dagger} = {\bf 1},$ where  ${\bf 1}$ is the identity operator. In terms of 
the corresponding matrix multiplication, this condition reads (${\bf I}$ is the $4\times 4$ identity matrix) 
\begin{equation}
\int_{(k'_z>0)} \frac{d^3K'}{(2\pi)^3} {\bf S}({\bf K}_1,{\bf K}')
{\bf S}({\bf K}_2,{\bf K}')^{\dagger}= (2\pi)^3 \delta^{(3)}({\bf K}_1 - {\bf K}_2)\, {\bf I}.
\end{equation}

We analyze in detail the scattering operator ${\bf S}_1$ corresponding to mirror M1 
in Fig.~1. 
We expand ${\bf S}_1$ up to second order of the deformation amplitudes 
 of the two lateral mirror's rough surfaces: 
\begin{equation}
{\bf S}_1={\bf S}_1^{(0)}+\delta{\bf S}_1^{(1)}+\delta{\bf S}_1^{(2)}.
\end{equation}
The scattering by ideal plane surfaces conserves $\omega$, ${\bf k}$ and polarization:
\begin{widetext}
\[
{\bf S}_1^{(0)}({\bf K},{\bf K}')= (2\pi)^3\delta^{(3)}({\bf K}-{\bf K}')
\left(
\begin{array}{cccc}
{\tilde t}_1^{\rm\scriptscriptstyle TE}({\bf k},\omega)&{ r}_1^{\rm\scriptscriptstyle TE}({\bf k},\omega) &0&0\\
{\tilde r}_1^{\rm\scriptscriptstyle TE}({\bf k},\omega)&{ t}_1^{\rm\scriptscriptstyle TE}({\bf k},\omega) &0&0\\
0&0&{\tilde t}_1^{\rm\scriptscriptstyle TM}({\bf k},\omega)&{ r}_1^{\rm\scriptscriptstyle TM}({\bf k},\omega)\\
0&0&{\tilde r}_1^{\rm\scriptscriptstyle TM}({\bf k},\omega)&{ t}_1^{\rm\scriptscriptstyle TM}({\bf k},\omega)\\
\end{array}
\right).
\]
 $r_1^p({\bf k},\omega)$ are the specular
 reflection 
coefficients as seen by the intracavity field presented in Sec.~III. 
The coefficients  ${\tilde r}_2^p({\bf k},\omega)$
play the same role for mirror M2.  Together they turn out to be the only relevant ones for the 
calculation of the Casimir effect, both in the ideal and rough cases. For simplicity, we have denoted 
 ${\tilde r}_2^p({\bf k},\omega)$
simply as ${r}_2^p({\bf k},\omega)$
everywhere in this paper, except in the present appendix.

The first  ($\ell=1$) and second-order  ($\ell=2$) corrections mix up different polarizations
and values of ${\bf k}$, but conserve the frequency:
\begin{equation}
\label{mdeltaS1}
\delta{\bf S}_1^{(\ell)}({\bf K},{\bf K}')= 2\pi \delta(k'_z-\sqrt{k^2-k'{}^2+k_z^2})
\left(
\begin{array}{cccc}
\delta{\tilde t}_1^{(\ell)}[{\rm \scriptstyle TE;TE}]&\delta r_1^{(\ell)}[{\rm \scriptstyle TE;TE}]&\delta{\tilde t}_1^{(\ell)}[{\rm \scriptstyle TE;TM}]&\delta r_1^{(\ell)}[{\rm \scriptstyle TE;TM}]\\
\delta{\tilde r}_1^{(\ell)}[{\rm \scriptstyle TE;TE}]&\delta t_1^{(\ell)}[{\rm \scriptstyle TE;TE}]&\delta{\tilde r}_1^{(\ell)}[{\rm \scriptstyle TE;TM}]&\delta t_1^{(\ell)}[{\rm \scriptstyle TE;TM}]\\
\delta{\tilde t}_1^{(\ell)}[{\rm \scriptstyle TM;TE}]&\delta r_1^{(\ell)}[{\rm \scriptstyle TM;TE}]&\delta{\tilde t}_1^{(\ell)}[{\rm \scriptstyle TM;TM}]&\delta r_1^{(\ell)}[{\rm \scriptstyle TM;TM}]\\
\delta{\tilde r}_1^{(\ell)}[{\rm \scriptstyle TM;TE}]&\delta t_1^{(\ell)}[{\rm \scriptstyle TM;TE}]&\delta{\tilde r}_1^{(\ell)}[{\rm \scriptstyle TM;TM}]&\delta t_1^{(\ell)}[{\rm \scriptstyle TM;TM}]\\
\end{array}
\right)
\end{equation}
\end{widetext}
All matrix elements above are functions of 
 frequency and of the initial and final momenta ${\bf k}'$ and ${\bf k}.$ 
The four elements $\delta r_1^{(\ell)}[p;p']$ are the matrix elements 
$\langle {\bf k},p|\, \delta{\cal R}_1^{(\ell)}\, |  {\bf k'},p'\rangle$
introduced in Sec.~III.

The transfer operators ${\bf T}$ provide the fields at the lefthand side of 
the mirror in terms of the fields at the righthand side. They can be obtained from the 
scattering operators as follows~\cite{lossy_cavities}: 
\begin{equation}\label{formal}
{\bf T}=-\left( {\bf P} _{-}-{\bf  S}\cdot{\bf P}_{+}\right)^{-1}\cdot
\left( {\bf P} _{+}-{\bf S}\cdot{\bf P} _{-}\right).
\end{equation}
The operators ${\bf P} _{+}$ and ${\bf P} _{-}$ are defined by
\[
{\bf P}_+({\bf K},{\bf K}')=(2\pi)^3 \delta^3({\bf K} - {\bf K}')
\left(
\begin{array}{cccc}
1&0&0&0\\
0&0&0&0\\
0&0&1&0\\
0&0&0&0\\
\end{array}
\right),
\]
\[
{\bf P}_- = {\bf 1} - {\bf P}_+.
\]

The intracavity field (see Fig.~1)
\[
{\bf e}^{\rm C}({\bf K}) =
\left(
\begin{array}{c}
\stackrel{\rightarrow}{e}_{\rm C}^{\rm\scriptscriptstyle TE}\\
\stackrel{\leftarrow}{e}_{\rm C}^{\rm\scriptscriptstyle TE}\\
\stackrel{\rightarrow}{e}_{\rm C}^{\rm\scriptscriptstyle TM}\\
\stackrel{\leftarrow}{e}_{\rm C}^{\rm\scriptscriptstyle TM}
\end{array}
\right)_{\bf K}
\]
\bigskip
\bigskip
\bigskip
\bigskip
\bigskip

is computed from 
\begin{equation}\label{opR1}
{\bf e}^{\rm C}({\bf K}) =
\int_{(k'_z>0)} \frac{d^3K'}{(2\pi)^3} {\bf R}({\bf K},{\bf K}'){\bf e}^{\rm in}({\bf K}'), 
\end{equation}
with 
\begin{equation}\label{opR2}
{\bf R} =  
{\bf T}_2\cdot {\bf P}_{+} \cdot {\bf S}_{\rm cav}
+{\bf T}_2\cdot {\bf P}_-,
\end{equation}
and where
${\bf S}_{\rm cav}$ is the scattering matrix for the cavity as a composed network.

The Casimir force 
on  mirror M1 is 
computed from the energy-momentum tensor component $T_{zz},$ 
evaluated at the intracavity region and at the outer side of 
the mirror, and averaged over the vacuum state:
\begin{equation}\label{Casimir}
F_{\rm PP} = \int d^2r\left[\langle T^{\rm L}_{zz}({\bf r})\rangle_{\mathrm{vac}}
-\langle T^{\rm C}_{zz}({\bf r})\rangle_{\mathrm{vac}}\right].
\end{equation}
We calculate $T^{\rm L}_{zz}({\bf r})$ using 
Eq.~(\ref{E_in})  and similar expansions for the magnetic field 
${\bf B}_{\rm L}^{\rm in}$
and for the outgoing fields: 
\begin{widetext} 
\begin{equation}\label{TzzVac}
\int d^2r\langle T^{\rm L}_{zz}({\bf r})\rangle_{\mathrm{vac}} = \frac{1}{2}
\int_{(k_z>0)}\frac{d^3 K}{(2\pi)^3}
\int_0^{\infty}\frac{d k'_z}{2\pi}
\hbar\omega
\cos^2\theta
\sum_p
\langle [e_{\rm L}^{\rm in}{}^p({\bf k},k_z) e_{\rm L}^{\rm in}{}^{p}({\bf k},k_z'){}^{\dagger}
+
e_{\rm L}^{\rm out}{}^p({\bf k},k_z) e_{\rm L}^{\rm out}{}^{p}({\bf k},k_z'){}^{\dagger} ]\rangle_{\mathrm{vac}},
\end{equation}
\end{widetext}
with $\cos\theta=k_z/K.$ A similar expression is found for the inner region
in terms of the intracavity fields. 

When taking the average over the vacuum state we use the commutation relation 
(\ref{commutation}) to find  
\begin{equation}\label{outside}
\langle e_{\rm L}^{\rm in}{}^p({\bf k},k_z) e_{\rm L}^{\rm in}{}^{p'}({\bf k},k_z'){}^{\dagger} \rangle_{\mathrm{vac}}=
2\pi\,A\, \delta(k_z-k_z').
\end{equation}
The outgoing field $e_{\rm L}^{\rm out}$ satisfies the same commutation relation, and provides an identical
contribution in Eq.~(\ref{TzzVac}). 
  
On the other hand, the commutation relation of the intracavity field 
is modified by the joint effect of the two mirrors. We derive the corresponding spectral density from 
Eqs.~(\ref{opR1}), (\ref{opR2}) and (\ref{outside}): 
\begin{equation}\label{inside}
\langle \stackrel{\rightarrow}{e}_{\rm C}^p({\bf k},k_z)\, 
 \stackrel{\rightarrow}{e}_{\rm C}^{p}({\bf k},k_z'){}^{\dagger} \rangle_{\mathrm{vac}}=
2\pi\,A\, g_p({\bf k}, \omega) \delta(k_z-k_z'),
\end{equation}
We obtain the explicit expressions for $\delta f_p^{(2\rm ii)}({\bf k}, \omega)$
and $\delta f_p^{(2\rm i)}({\bf k}, \omega)$ given by 
Eqs.~(\ref{fii_explicit}) and (\ref{fi_explicit}) with the help of a computer algebra system.
Remarkably, they only contain  the rough reflection coefficients associated to internal reflections
[only 4 out of 16 elements in Eq.~(\ref{mdeltaS1}) for instance]. 

The spectral density for $\stackrel{\leftarrow}{e}_{\rm C}$ is modified, with respect to
the free-space case, by the same generalized Airy function $g_p({\bf k}, \omega),$
as far as second order terms containing rough reflections at the {\it same} mirror are concerned.
After replacing (\ref{outside}) into (\ref{TzzVac}), and 
using (\ref{Casimir}) and (\ref{inside}), we derive the result given by
Eq.~(\ref{Fcas}) for the Casimir force.

\section{Proximity Force Approximation as a limiting  case}\label{AppB}

In 
this appendix, we derive the PFA result for the energy correction as a limiting case
of the general results of Sec.~III. 
This will bring a deeper understanding of the PFA, by showing its connection with 
some specific properties of the reflection operators. These properties must be satisfied regardless
of the particular model considered for the material medium, and  are
related to the specular reflection by displaced plane mirrors (specular limit).  

In the PFA regime, the Fourier profile functions $H_j(\Delta{\bf k})$ are sharply peaked around
$\Delta {\bf k}={\bf 0}.$ Thus, we may replace ${\bf k'}$ by ${\bf k}$ in the argument
 of the non-specular
coefficients appearing in the r.-h.-s. of (\ref{R2}) and (\ref{R1}). The resulting expressions correspond to 
the case of ideal plane mirrors which are displaced from the associated reference planes at $z=0$ and $z=L.$
In this case, the reflection is modified with respect to the non-perturbed case just by the effect of the 
propagation from the reference plane to 
the mirror and back. Up to second order, this amounts to take the 
series expansion of the exponential factor
\[
e^{-2\kappa h_j}\approx 1-2\kappa h_j +2 \kappa^2 h_j^2,
\]  
yielding 
\begin{equation}\label{R1PFA}
R_{j;pp'}^{(1)}({\bf k},{\bf k};\xi) = -2\kappa\, r_j^p({\bf k},\xi)\delta_{p,p'}
\end{equation}
and 
\begin{equation}\label{R2PFA}
R_{j;p}^{(2)}({\bf k},{\bf k};\xi)
 = 2\kappa^2 r_j^p({\bf k},\xi).
\end{equation}
Eqs.~(\ref{R1PFA}) and (\ref{R2PFA}) are
general properties of the reflection coefficients, and are useful for checking  
explicit  calculations, regardless of the specific model considered for the material medium.
When using these results to compute $G({\bf 0})$ from Eqs.~(\ref{G})-(\ref{G_ii}), we obtain
Eq.~(\ref{PFA2}), with the Casimir energy in the ideal case  given by
(\ref{plane-plane}). This verifies the PFA limit as discussed in the end of Sec.~III.

\section{ Perfect mirrors }

In Sec.~V, the perfectly-reflecting limit was derived from the plasma model results by taking 
$\lambda_P \ll k^{-1}, L.$ 
This appendix presents an alternative, simpler derivation, in which the usual model of perfect reflectors 
is taken from the start. 
The case of corrugation along a fixed direction in the  $xy$ plane
(say the direction along the $x$-axis, with $h_1(x,y)=h_1(x)$)  was considered by 
Ref.~\cite{Emig}. In this case, 
the calculation can be considerably simplified by taking a convenient definition for the 
field polarizations, which then turn out to be not coupled by the scattering from 
the surface.
On the other hand, in this paper we consider arbitrary small-amplitude 
deformations, so that the coupling between different polarizations has to be 
taken into account. 

For the mirror near $z=0,$ we take the
boundary condition 
\begin{equation}\label{BCperf}
{\hat n}_1(x,y) \times {\bf E}(x,y,h_1(x,y))={\bf 0},
\end{equation}
where 
${\hat n}_1(x,y)$ is the unitary vector normal to the tangent plane at the point
$(x,y):$ 
\[
{\hat n}_1(x,y)=\frac{{\hat z}-{\bf \nabla} h_1}{\sqrt{1+({\bf \nabla} h_1)^2}}.
\] 
As explained in Sec.~II, positive values of $h_1(x,y)$ are defined  along the positive
$z$ axis, and the (intracavity) incident field, as given 
by (\ref{incident})
 propagates along the negative $z$ axis with 
a phase factor $e^{-ik_z z}.$ 

Solving (\ref{BCperf}) up to first 
order of $h_1(x,y),$ we determine the complete first-order reflection operator.
Using the matrix notation introduced 
in Sec.~IV, the first-order non-specular coefficients
defined by Eq.~(\ref{R1}) are 
written as
\begin{equation}
\label{R1perf}
{\bf R^{(1)}}({\bf k},{\bf k}';\omega) = 2i\left(
\begin{array}
[c]{cc}%
k_z' C
 & \omega S/c
\\
\frac{\omega k_z'}{c k_z} S & \frac{k k'}{k_z}- \frac{\omega^2}{c k_z} C
\end{array}
\right),
\end{equation}
where $C$ and $S$ are the cosine and sine of the angle between ${\bf k}$ and ${\bf k}'.$

We only need the 
diagonal elements of the second-order reflection operator, which we collect from 
the  solution of  (\ref{BCperf}) up to second order. The corresponding non-specular coefficients 
[see Eq.~(\ref{R2})] are 
\begin{equation}\label{2tete}
 R^{(2)}_{\scriptscriptstyle \rm TE}({\bf k},{\bf k}';\omega)= 
2 k_zk_z' C^2
+2\frac{\omega^2 k_z}{c^2 k_z'} S^2
\end{equation}
\begin{equation}\label{2tmtm}
 R^{(2)}_{\scriptscriptstyle\rm TM}({\bf k},{\bf k}';\omega)= 
 -\frac{2}{k_z k_z'}
\left(k k'- \omega^2C/c^2\right)^2
-2\frac{\omega^2 k_z'}{c^2 k_z} S^2
\end{equation}

These results satisfy the specular limit discussed in Appendix B. When replaced into the general
expressions of Sec.~III, they reproduce the results for the perfectly-reflecting limit 
of Sec.~V.


\begin{thebibliography}{99}
\bibitem{foot1} Note that the matrix elements of the reflection operators have  dimension of 
area, as can be checked from Eq.~(\ref{reflection-operator-explicit}),
 so that the loop function as given by (\ref{fii_explicit}) is dimensionless. Moreover,
the diagonal elements given by (\ref{R2}) are actually proportional to $A$
because of Eq.~(\ref{area2}).
Hence the loop function is independent of $A$ as expected.
\end{thebibliography}
\end{document}